\documentclass[aps,showpacs,preprintnumbers,amsmath,amssymb,twocolumn,superscriptaddress,floatfix,nofootinbib,10pt,pre]{revtex4-1}

\usepackage[linesnumbered,lined,boxed,commentsnumbered,ruled,vlined]{algorithm2e}
\usepackage{algpseudocode}
\usepackage{amssymb, amsmath, amsfonts}
\usepackage{amsthm, mathrsfs, amsopn}
\usepackage{bm,bbm}
\usepackage{booktabs}
\usepackage{caption}
\usepackage{centernot}
\usepackage{colortbl}
\usepackage{comment}
\usepackage{dcolumn}
\usepackage{epstopdf}
\usepackage{float}
\usepackage{graphicx}
\usepackage[bottom]{footmisc}
\usepackage{mathrsfs}
\usepackage{mathtools}
\usepackage{multirow}
\usepackage{ragged2e}
\usepackage{subfigure}
\usepackage{tabularx}
\usepackage{tikz}
\usepackage{verbatim}
\usepackage{hyperref}
\usepackage{cleveref}
\usepackage{amsthm}

\DeclareCaptionJustification{justified}{\justifying}
\hypersetup{colorlinks=true, citecolor=orange, urlcolor=blue, linkcolor=magenta}

\begin{document}

%\title{The $d$-dimensional Kuramoto model: from networks to matrix-weighted networks}

\title{Synchronization of higher-dimensional Kuramoto oscillators on networks: from scalar to matrix-weighted couplings}

\author{Anna Gallo}
\affiliation{Centro Studi e Ricerche `Enrico Fermi' (CREF), Via Panisperna 89A, 00184 Rome (Italy)}
\affiliation{IMT School for Advanced Studies, Piazza San Francesco 19, 55100 Lucca (Italy)}
\affiliation{Institute for Applied Computing `Mauro Picone' (IAC), National Research Council, Via Madonna del Piano 10, 50019 Sesto Fiorentino (FI) (Italy)}

\author{Renaud Lambiotte}
\email{renaud.lambiotte@maths.ox.ac.uk}
\affiliation{Mathematical Institute, University of Oxford, Woodstock Rd, Oxford OX2 6GG (United Kingdom)}

\author{Timoteo Carletti}
%\email{timoteo.carletti@unamur.be}
\affiliation{Department of mathematics and Namur Institute for Complex Systems, naXys\\University of Namur, Rue Graf\'e 2, B5000 Namur (Belgium)}

\date{\today}

\begin{abstract}
The Kuramoto model is the paradigmatic model to study synchronization in coupled oscillator systems. In its classical formulation, the oscillators move on the unit circle, each characterized by a scalar phase and a natural frequency, by interacting through a sinusoidal coupling. In this work, we propose a $d$--dimensional generalization in which oscillators are represented as unit vectors on the $(d-1)$--sphere and interact through a matrix-weighted network (MWN), a recently introduced framework where links are endowed with a matrix weight instead of a scalar one. We derive necessary conditions for global synchronization via a Master Stability Function approach: the existence of a synchronous solution requires identical frequency matrices across nodes and, in the MWN case, a coherence condition on the network structure. Through a suitable change of variables, the stability analysis reduces the full $Nd$--dimensional problem to a family of $d$--dimensional eigenvalue problems, each one parametrized by the eigenvalue of a suitable scalar weighted Laplacian, showing that the synchronous solution is locally stable for any positive coupling strength $K$ on any connected network. Analytical results are complemented by numerical simulations.
\end{abstract}

% \pacs{}

\maketitle

\section{Introduction}
\label{sec:intro}

Networks offer a natural language for modelling complex systems across disciplines, from sociology and economics to biology and technology~\cite{newman2018networks}. In many such systems, the interplay between network structure and dynamical behaviour is fundamental. In linear settings, this relationship is largely mediated by the spectra of graph-associated operators (e.g., the Laplacian), which often govern stability, relaxation, and diffusion-like processes~\cite{porter2016dynamical}. Nonlinear dynamics on networks is considerably less tractable, typically demanding model-specific analytical tools and approximations~\cite{moreno2002epidemic,chandra2019complexity}. Among nonlinear phenomena, synchronization occupies a central position: it denotes the spontaneous emergence of coherent behaviour among interacting dynamical units~\cite{strogatz2003synchronization,osipov2007synchronization,arenas2008synchronization,ghosh2023measure}. Synchronization is observed in systems as diverse as power grids, neural populations, and coordinated social dynamics~\cite{vallacher2005dynamics,motter2019spontaneous,muller2022neural}.

A paradigmatic framework for studying synchronization is the Kuramoto model. Since its introduction in 1975~\cite{kuramoto1975international}, it has become a canonical model for collective phase dynamics in coupled oscillator systems~\cite{acebron2005kuramoto,ADGKMZ2008}. In its classical form, the model describes $N$ oscillators on the unit circle, each characterized by a phase $\theta_i \in [0,2\pi)$ and a natural frequency $\omega_i$. 
A positive coupling between oscillators tends to align their phases, leading to collective synchronization above a critical coupling strength. Mathematically, the evolution of each oscillator reads
\begin{align}
\label{eq:Kmodel}
\frac{d{\theta}_i}{dt} = \omega_i + \frac{K}{N}\sum_{j} A_{ij} \sin(\theta_j - \theta_i),
\end{align}
where $K>0$ is the coupling strength and $A_{ij}$ is the adjacency matrix encoding the interaction network, i.e., $A_{ij}=1$ if and only if nodes $i$ and $j$ are connected, and $0$ otherwise. Owing to its simplicity and broad relevance in physics, biology, engineering, and the social sciences, the Kuramoto model has served as a prototype for probing how network architecture shapes nonlinear oscillatory dynamics \cite{jadbabaie2004stability,coutinho2013kuramoto,rodrigues2016kuramoto,lopes2016synchronization,li2017optimizing}.

Most studies assume scalar, typically positive, coupling weights. More recently, however,  works have replaced the scalar coupling with a \emph{coupling matrix} acting on two-dimensional oscillator representations~\cite{buzanello2022matrix,de2023generalized,fariello2024exploring}. In the resulting \textit{frustrated} Kuramoto model, matrix coupling can induce forms of frustrated synchronization that have no direct analogue in scalar-coupled systems. 
%Existing formulations generally take the coupling matrix to be uniform across nodes. In this work, we relax this assumption and allow the coupling to depend on node indices, as we detail below.

% Over the past two decades, researchers have explored several distinct directions for generalizing the Kuramoto model beyond the circle. One natural extension replaces the circle with higher-dimensional unit spheres, where oscillator states are represented as unit vectors in $\mathbb{R}^d$ rather than angles. In~\cite{tanaka2014solvable}, Tanaka extends the Kuramoto model to describe particles that interact on a high-dimensional unit sphere and that remain on it without requiring additional constraints, as in~\cite{ritort1998solvable}. Moreover, in such an extension, the dynamic variables are vectors rather than matrices, as in~\cite{gu2007phase}. More recently, the Kuramoto model has been studied on spheres in~\cite{crnkic2021synchronization} and~\cite{lipton2021kuramoto}. In the former, the authors prove that the phenomenon of synchronization depends on the dimension of the sphere, with synchronization occurring faster on higher-dimensional spheres. In the latter, Lipton and coauthors~\cite{lipton2021kuramoto} show that the long-term dynamics converge to a low-dimensional invariant manifold.

Over the past two decades, researchers have also explored generalisations of the Kuramoto model to higher dimensions. A natural extension replaces the circle---that is, the unit $1$--sphere---with the unit $(d-1)$--sphere, $\mathbb{S}^{d-1}\subset\mathbb{R}^d$, $d\geq 2$, so that oscillator states are unit vectors rather than phases~\cite{ritort1998solvable,gu2007phase,tanaka2014solvable,crnkic2021synchronization,lipton2021kuramoto}. In this higher-dimensional setting, each oscillator is represented by a unit vector rather than a phase, and the coupling term is replaced by the projection of neighbouring vectors onto the local tangent plane of the sphere, ensuring that the dynamics remain confined to $\mathbb{S}^{d-1}$. Concretely,~\eqref{eq:Kmodel} is replaced by
\begin{align}
    \label{eq:Kdmodel}
    \frac{d{\vec{x}}_i}{dt} = \Omega_i\vec{x}_i+\frac{K}{N}\sum_{j}A_{ij} \left( \vec{x}_j-\langle \vec{x}_j,\vec{x}_i\rangle \vec{x}_i\right)\, ,
\end{align}
where $\vec{x}_i\in\mathbb{S}^{d-1}$, $\Omega_i$ are $(d\times d)$ antisymmetric matrices and $\langle\cdot,\cdot\rangle$ denotes the scalar product in $\mathbb{R}^d$. The coupling term $\left( \vec{x}_j-\langle \vec{x}_j,\vec{x}_i\rangle \vec{x}_i\right)$
is the projection of $\vec{x}_j$ onto the tangent plane of the sphere
at~$\vec{x}_i$, which ensures that the dynamics preserve the set
$\|\vec{x}_i\|=1$. Note that when all oscillators share the same intrinsic frequency (e.g., $\Omega_i\equiv \Omega$), the dynamics reduces---in a rotating frame---to a consensus-type evolution for $d$--dimensional vectors,
\begin{align}
    \frac{d{\vec{x}}_i}{dt}=\sum_j A_{ij}(\vec{x}_j-\vec{x}_i),
\end{align}
where the dynamics is constrained on the sphere via Lagrange multipliers \cite{OlfatiSaber2006}.

Most studies of the higher-dimensional Kuramoto model focus on the all-to-all (mean-field) case. In that regime, the nature of the synchronization transition depends on the parity of the dimension~\cite{chandra2019continuous}: for odd $d$, synchronization emerges discontinuously as the coupling increases from zero, whereas for even $d$ the transition is continuous and occurs at a positive critical coupling. Comparatively fewer works address the role of network connectivity, where each oscillator interacts only with a subset of neighbours. In this setting, stability of the synchronous solution has been established by using Lyapunov methods and, in particular, the \textit{LaSalle invariance principle}~\cite{la1966invariance,shi2023synchronization}. Notably,~\cite{markdahl2021almost} proves almost global convergence to synchronization on networks over the $d$--sphere, showing that for even $d$, phase-locking is attained even in the presence of non-identical natural frequencies.

Lohe~\cite{lohe2009non} introduced non-Abelian extensions defined on compact Lie groups, showing that synchronization properties of the classical model carry over to this richer algebraic setting. More recently, in~\cite{ha2022emergent}, Ha and co-authors studied emergent synchronization behaviour for Kuramoto-type models on Stiefel manifolds, which unify several previously considered geometries by including spheres and orthogonal groups.  All the aforementioned works and others highlight the broad applicability of Kuramoto-type dynamics beyond the unit circle.

The goal of this paper is to extend the work of Lipton \textit{et al.}~\cite{lipton2021kuramoto} by studying the $d$--dimensional Kuramoto model on arbitrary network topologies, with the aim of clarifying how connectivity shapes synchronization on higher-dimensional spheres. Our approach builds on the framework of matrix-weighted networks (MWNs)~\cite{tian2025matrix,gallo2025global}, in which interactions are not described by scalar coupling strengths but by orthonormal matrices that transform the exchanged vectors. This structure is particularly well suited to modelling the mixing of high-dimensional signals between neighbouring nodes. Within this setting, we derive necessary conditions for the existence of globally synchronized states and investigate their stability via Master Stability Function (MSF) techniques.

More precisely, to prove the emergence of global synchronization we exploit the coherence assumption and the commuting property of the antisymmetric matrix $\Omega$ with the matrix weights, $\mathbf{R}_{ij}$ (see Section~\ref{sec:mwns}) to perform a change of coordinates allowing us to get rid of the rotation matrices and so to map the Kuramoto model defined on MWN onto a model defined on weighted scalar network. We then perform a linear stability
analysis on the latter model in a neighbourhood of a periodic reference solution; the existence of such solution is again guaranteed by the coherence condition. In this way we are able to introduce a suitable Laplace matrix whose spectrum allows to conclude about the stability of the synchronous solution. Hence coherence is a necessary
ingredient for existence of the synchronization that together with the property of commutation, $[\mathbf{R}_{ij},\Omega]=0$, determine an interesting interplay between the MWN structure, e.g., its matrix
weights, and the dynamical system.

The paper is organized as follows. In Section~\ref{sec:dKur}, we introduce the $d$--dimensional Kuramoto model on complex networks, providing the analytical analysis of the emergence of global synchronization in such a framework in subsection~\ref{sec:sync}. Whereafter, in Section~\ref{sec:mwns}, we define the $d$--dimensional Kuramoto model on matrix-weighted networks and derive the necessary conditions for the existence of a synchronous solution via a Master Stability Function approach.  Finally, in Section~\ref{sec:conc}, we summarize our findings and discuss open problems and future perspectives.

\section{The $d$--dimensional Kuramoto model on complex networks}
\label{sec:dKur}

\subsection{Defining the model}

We consider oscillators moving on the unit $(d-1)$--dimensional sphere, $d\geq 2$, whose positions can be defined by the vectors $\vec{x}_i\in \mathbb{S}^{d-1}\subset \mathbb{R}^d$, $i=1,\dots,N$.
We encode the structure of the complex network by the $N\times N$ adjacency matrix, $\mathbf{A}$, such that $A_{ij}=1$ if and only if oscillators (nodes) $i$ and $j$ are connected, and $0$ otherwise. We also define $k_i=\sum_j A_{ij}$ as the degree of node $i$.
We consider the dynamical system~\eqref{eq:Kdmodel}, and rewrite it in the equivalent form
\begin{align}
\label{eq:dKurNet}
\frac{d{\vec{x}}_i}{dt} = \Omega_i \vec{x}_i + Z_i(\vec{x})-\langle Z_i(\vec{x}),\vec{x}_i\rangle \vec{x}_i\,,
\end{align}
where $\Omega_i$ is an antisymmetric $(d\times d)$ matrix, and $\vec{x}=(\vec{x}^\top_1,\dots,\vec{x}^\top_N)^\top$ denotes the stacked state of the network. The term $Z_i(\vec{x})\in\mathbb{R}^d$ represents the \emph{local mean field} acting on node~$i$, i.e.\ the aggregate influence of its neighbours, and is defined by
\begin{align}\label{eq:defZ}
Z_i(\vec{x})=\frac{K}{N}\sum_j A_{ij}\vec{x}_j.
\end{align}

Let us first verify that the dynamics induced by
Eq.~\eqref{eq:dKurNet} preserves the sphere, namely that if the
initial conditions lie on the sphere, the flow remains on the
sphere for all time.  We compute the time derivative of the
squared norm $\langle \vec{x}_i,\vec{x}_i\rangle$ of the $i$-th
oscillator:
\begin{align}
\label{eq:belongsph}
\frac{d}{dt}\langle &\vec{x}_i,\vec{x}_i\rangle = 2 \Big\langle \vec{x}_i,\frac{d{\vec{x}}_i}{dt}\Big\rangle\nonumber\\
&= 2 \langle \vec{x}_i, \Omega_i \vec{x}_i +Z_i(\vec{x})-\langle Z_i(\vec{x}),\vec{x}_i\rangle \vec{x}_i\rangle\nonumber\\
&= 2 \langle \vec{x}_i, \Omega_i \vec{x}_i \rangle +2 \langle \vec{x}_i,Z_i(\vec{x})\rangle - 2\langle Z_i(\vec{x}),\vec{x}_i\rangle \langle \vec{x}_i,\vec{x}_i\rangle\nonumber\\
&=2 \langle \vec{x}_i, \Omega_i \vec{x}_i \rangle + 2\langle \vec{x}_i,Z_i(\vec{x})\rangle (1-\langle \vec{x}_i,\vec{x}_i\rangle)=0\, ,
\end{align}
where we used the antisymmetry of $\Omega_i$ to conclude
$\langle \vec{x}_i, \Omega_i \vec{x}_i \rangle = 0$, while the
remaining term vanishes because
$\langle \vec{x}_i,\vec{x}_i\rangle = 1$.  Since the derivative
is zero whenever $\|\vec{x}_i\| = 1$, the condition
$\langle \vec{x}_i,\vec{x}_i\rangle = 1$ is preserved for all
$t > 0$ provided it holds at $t = 0$.

\subsection{Global synchronization}
\label{sec:sync}

We first determine the conditions under which system~\eqref{eq:dKurNet} exhibits global synchronization, that is the scenario in which all oscillators follow the same behaviour at unison, $\vec{x}_i(t)=\vec{s}(t)\in\mathbb{S}^{d-1}$ for all $i$. Such a solution can exist only if the frequency matrices $\Omega_i$ are identical across nodes. We therefore assume $\Omega_i=\Omega$ for all $i=1,\dots,N$, where $\Omega$ is a fixed antisymmetric $(d\times d)$ matrix.

Let us first show in detail how to reduce the problem to a consensus dynamics constrained to the unit sphere.
Under the assumption of identical matrices, $\Omega_i\equiv\Omega$, for all $i=1,\dots,N$,  system~\eqref{eq:dKurNet} can be simplified by changing variables and passing to a rotating reference frame. In this way, the resulting system reduces to a ``pure'' consensus problem on the sphere.

 To show so, we define the change of variables
\begin{align}
    \vec{\xi}_i=e^{-\Omega t}\vec{x}_i\quad \forall\:i=1,\dots, N\, ,
\end{align}
so that  equation~\eqref{eq:dKurNet} rewrites for all $i$
\begin{align}
\label{eq:dKurNetb}
\frac{d{\vec{\xi}}_i}{dt} &= -\Omega \vec{\xi}_i +e^{-\Omega t}\frac{d{\vec{x}}_i}{dt}\\
&= -\Omega \vec{\xi}_i+e^{-\Omega t}\Omega \vec{x}_i+e^{-\Omega t}Z_i(\vec{x})-\langle  Z_i(\vec{x}),\vec{x}_i\rangle e^{-\Omega t}\vec{x}_i\,.\nonumber
\end{align}

Since $\Omega$ commutes with $e^{-\Omega t}$, the first two terms cancel. By direct computation, one shows that
\begin{align}
   e^{-\Omega t}Z_i(\vec{x}) &= e^{-\Omega t}\sum_j A_{ij} \vec{x}_j=\sum_j A_{ij} e^{-\Omega t}\vec{x}_j\nonumber\\
   &=\sum_j A_{ij} \vec{\xi}_j=Z_i(\vec{\xi})\, ,
\end{align}
and also
\begin{align}
    \langle  Z_i(\vec{x}),\vec{x}_i\rangle &= \sum_j A_{ij} \langle  \vec{x}_j,\vec{x}_i\rangle=\sum_j A_{ij} \langle  \vec{x}_j,e^{\Omega t}e^{-\Omega t}\vec{x}_i\rangle\nonumber\\
   &=\sum_j A_{ij} \langle  e^{-\Omega t}\vec{x}_j,e^{-\Omega t}\vec{x}_i\rangle=\sum_j A_{ij} \langle  \vec{\xi}_j,\vec{\xi}_i\rangle\nonumber\\
   &=\langle  Z_i(\vec{\xi}),\vec{\xi}_i\rangle\, ,
\end{align}
where we used the fact that $\Omega^\top = -\Omega$.

In the rotating frame, the dynamics therefore take the form
\begin{align}
\label{eq:dKurNetc}
\frac{d{\vec{\xi}}_i}{dt} = Z_i(\vec{\xi})-\langle  Z_i(\vec{\xi}),\vec{\xi}_i\rangle \vec{\xi}_i\, , \quad \forall\:i=1,\dots,N\, .
\end{align}
We look for a reference solution $\vec{x}_i=\vec{s}$ of~\eqref{eq:dKurNet}, with $\Omega_i=\Omega$ for all $i$, and obtain
\begin{align}
\label{eq:dKurNetsync}
\frac{d{\vec{s}}}{dt} = \Omega \vec{s} +Z_i(\vec{s})-\langle Z_i(\vec{s}),\vec{s}\rangle \vec{s}\, . 
\end{align}
By defining $Z_i$ as in Eq.~\eqref{eq:defZ}, we obtain
\begin{align}
 Z_i(\vec{s})=\frac{K}{N}\sum_j A_{ij}\vec{s}=\frac{K}{N}k_i\vec{s}\, ,
\end{align}
thus~\eqref{eq:dKurNetsync} rewrites
\begin{align}
\label{eq:dKurNetsync2}
\frac{d{\vec{s}}}{dt} &= \Omega \vec{s} +\frac{K}{N}k_i\vec{s}-\frac{K}{N}k_i\langle \vec{s},\vec{s}\rangle \vec{s}\nonumber\\
   &=\Omega \vec{s} +\frac{K}{N}k_i\vec{s}-\frac{K}{N}k_i\langle \vec{s},\vec{s}\rangle \vec{s}=\Omega \vec{s}\, ,
\end{align}
where we used the fact that $\langle \vec{s},\vec{s}\rangle=1$. The solution of~\eqref{eq:dKurNetsync2} is then given by
\begin{align}
\vec{s}(t)=e^{\Omega t}\vec{s}(0)\, ,
\end{align}
and, since $\Omega$ is antisymmetric, the latter is simply a uniform rotation on the sphere. Observe that for more general function $Z_i$, Eq.~\eqref{eq:dKurNetsync} could not be satisfied unless we impose the network to be regular, as the right hand side depends on $i$ while the left hand side does not.

%\subsection{Linear stability}
We are now interested in the stability of the synchronous solution $\vec{x}_i(t)=\vec{s}(t)$, for all $i$. To this end, we introduce the perturbations $\delta\vec{x}_i=\vec{x}_i-\vec{s}$ and study their evolution. 
Since both $\vec{x}_i$ and $\vec{s}$ lie on the sphere, $\delta\vec{x}_i$ is, to first order, orthogonal to $\vec{s}$; in other words, the perturbation is tangent to the sphere at $\vec{s}$.
This can be seen by computing
\begin{align}
1 &= \langle \vec{x}_i,\vec{x}_i\rangle = \langle \delta\vec{x}_i+\vec{s},\delta\vec{x}_i+\vec{s}\rangle\nonumber\\
   &=\langle \vec{s},\vec{s}\rangle+2\langle \vec{s},\delta\vec{x}_i\rangle+\langle \delta\vec{x}_i,\delta\vec{x}_i\rangle\nonumber\\
   &=1+2\langle \vec{s},\delta\vec{x}_i\rangle+\mathcal{O}(|\delta\vec{x}_i|^2)\,.
\end{align}

To derive the evolution equation for $\delta\vec{x}_i$, we substitute $\vec{x}_i=\vec{s}+\delta\vec{x}_i$ into~\eqref{eq:dKurNet}, use~\eqref{eq:dKurNetsync2}, and retain only terms that are linear in $\delta\vec{x}_i$. In this way, for each $i=1,\dots,N$ we obtain:
\begin{align}
\label{eq:dKurNetlin}
\frac{d{\delta\vec{x}}_i}{dt} = \Omega \delta\vec{x}_i &+\sum_\ell \frac{\partial Z_i}{\partial \vec{x}_\ell}(\vec{s})\delta\vec{x}_\ell-\langle Z_i(\vec{s}),\vec{s}\rangle \delta \vec{x}_i-\langle Z_i(\vec{s}),\delta \vec{x}_i\rangle\vec{s}\, .
\end{align}
Assume again to deal with $Z_i$ defined as in Eq.~\eqref{eq:defZ}. Thus $Z_i(\vec{s})=\frac{K}{N}k_i\vec{s}$, and, recalling that $\langle \vec{s}, \delta\vec{x}_i\rangle = 0$, Eq.~\eqref{eq:dKurNetlin} reduces to
\begin{align}
\label{eq:dKurNetlin2}
\frac{d{\delta\vec{x}}_i}{dt} &= \Omega \delta\vec{x}_i +\frac{K}{N}\sum_\ell A_{i\ell}\delta\vec{x}_\ell-\frac{K}{N}k_i\delta\vec{x}_i\nonumber\\
&= \Omega \delta\vec{x}_i -\frac{K}{N}\sum_\ell L_{i\ell}\delta\vec{x}_\ell \quad \forall\:i=1,\dots,N\,,
\end{align}
where we have used the Laplace matrix $L_{i\ell}=k_i\delta_{i\ell}-A_{i\ell}$.

By introducing the stack $(Nd)$--vector $\delta\vec{x}=(\delta\vec{x}^\top_1,\dots,\delta\vec{x}^\top_N)^\top$, and by using the Kronecker product, we can rewrite Eq.~\eqref{eq:dKurNetlin2} in compact form as follows
\begin{align}
\label{eq:dKurNetlin3}
\frac{d{\delta\vec{x}}}{dt}=\left(\mathbf{I}_N\otimes \Omega-\frac{K}{N} \mathbf{L}\otimes \mathbf{I}_d\right)\delta \vec{x}\, .
\end{align}
To reduce the dimensionality of the latter system and thus make analytical progress, we invoke the orthonormal eigenbasis $\phi^{(\alpha)}\in\mathbb{R}^N$ of the symmetric Laplace matrix $\mathbf{L}$, and the associated eigenvalues, $\Lambda^{(\alpha)}\geq 0$, $\alpha=1,\dots, N$. In order to study the stability of Eq.~\eqref{eq:dKurNetlin3}, we project $\delta\vec{x}_i$ onto the eigenbasis
\begin{align}
\label{eq:projectxiphi}
\delta\vec{x}_i=\sum_{\alpha}\delta\hat{x}_\alpha \phi^{(\alpha)}_i\, ,
\end{align}
where $\phi^{(\alpha)}_i\in\mathbb{R}$ is the $i$-th scalar component of the $\alpha$-th 
eigenvector $\phi^{(\alpha)}$ of the $N\times N$ Laplace matrix $\mathbf{L}$.
% , and substitute into~\eqref{eq:dKurNetlin3}. Thus, we obtain
Substituting \eqref{eq:projectxiphi} into \eqref{eq:dKurNetlin3}, and by using the eigenvalue relation $\sum_\ell L_{i\ell}\phi^{(\alpha)}_\ell = \Lambda^{(\alpha)}\phi^{(\alpha)}_i$, we multiply both sides by $\phi^{(\alpha)}_i$ and sum over $i$, exploiting 
orthonormality of the eigenbasis, to obtain
\begin{align}
\label{eq:dKurNetlin3proj}
\frac{d{\delta\hat{x}}_{\alpha}}{dt}=\left[\Omega-\frac{K}{N} \Lambda^{(\alpha)}\mathbf{I}_d\right]\delta \hat{x}_{\alpha}\,,
\end{align}
that holds true for each mode $\alpha$ independently.
For each $\alpha$, we compute the $d$ eigenvalues $\lambda_j$, depending on $\Lambda^{(\alpha)}$, of the characteristic equation
\begin{align}
\label{eq:MSF1}
0=\det \left(\Omega-\frac{K}{N}\Lambda^{(\alpha)}\mathbf{I}_d-\lambda \mathbf{I}_d\right)\,.
\end{align}
It follows that if
\begin{align}
\label{eq:sync}
\max_{\alpha}\rho_{\alpha}(\Lambda^{(\alpha)}) < 0\, ,
\end{align}
where $\rho_{\alpha} = \max_{j}\left[\Re\lambda_j(\Lambda^{(\alpha)})\right]$, then we can conclude that the system synchronizes. Indeed for $\alpha>1$, $\delta\hat{x}_\alpha(t)\rightarrow 0$. Let us observe that $\frac{d\delta \hat{x}_1}{dt}=\Omega\delta\hat{x}_1$, whose solution is again a rotation by $\Omega$, $\hat{x}_1(t)=e^{\Omega t}\hat{x}_1(0)$. We can thus conclude that for large $t>0$, we have $\delta\vec{x}_i(t)\sim e^{\Omega t}\hat{x}_1(0)/\sqrt{N}=e^{\Omega t}\sum_j \delta\vec{x}_j(0)/N$, where we used the projection onto the eigenbasis and $\phi^{(1)}=(1,\dots,1)^\top/\sqrt{N}$. Eventually, we obtain $\vec{x}_i(t)\sim e^{\Omega t}(\vec{s}(0)+\sum_j \delta\vec{x}_j(0)/N$ for $t\rightarrow \infty$.

Let us observe that Eq.~\eqref{eq:sync} always holds true; indeed because $\Omega$ is antisymmetric, its eigenvalues are pure imaginary or zero numbers, hence $\rho_{1} = 0$ and $\rho_{\alpha} = -\frac{K}{N}\Lambda^{(\alpha)}<0$ for $\alpha=2,\dots,N$, being $K>0$.

\section{The $d$--dimensional Kuramoto model on matrix-weighted networks}
\label{sec:mwns}

\subsection{From scalar to matricial interactions}

In the previous section, coupling between nodes was mediated by a scalar weight, so that interactions acted identically and independently on each coordinate. Here we consider the more general situation in which each edge is endowed with a linear transformation that reshapes the exchanged high-dimensional signal. To this end, we couple the $N$ oscillators through a \emph{Matrix-Weighted Network}~\cite{tian2025matrix} (MWN), where to any existing link connecting nodes $i$ and $j$, we associate the weight matrix $\mathbf{W}_{ij}\in \mathbb{R}^{d\times d}$. For any $i,j$, such a matrix is defined as $\mathbf{W}_{ij}=w_{ij}\mathbf{R}_{ij}$, where $w_{ij}$ is a scalar positive weight and $\mathbf{R}_{ij}$ is an orthonormal transformation matrix, $||\mathbf{R}_{ij}||_2=1$. In the following, for simplicity, we will assume the transformation to be a rotation.

Here, we assume the MWN to be coherent in the sense of~\cite{tian2025matrix}, namely for any oriented cycle, the product of the associated rotation matrices equals the identity. More formally, let us recall that an 
\textit{oriented cycle} in a graph is a closed sequence of edges $e_0=(i_0,i_1)\to e_1=(i_1,i_2)\to\cdots\to e_k=(i_k,i_0)$, and the aforementioned coherence condition requires that for any such cycle, the ordered product of the associated rotation matrices satisfies
\begin{align}
    \mathbf{R}_{i_0 i_1}\mathbf{R}_{i_1 i_2}\cdots
    \mathbf{R}_{i_k i_0}=\mathbf{I}_d\,.
\end{align}

By Proposition 1 of~\cite{gallo2026turing}, this is equivalent to the existence of orthonormal matrices $\mathbf Q_1,\dots,\mathbf Q_N\in O(d)$ such that $\mathbf R_{ij}=\mathbf Q_i^\top\mathbf Q_j$ for every edge $(i,j)$. The coherence property allows us to define, for each node $i$, the transformation matrix $\mathbf O_{1i}$ as the product of the link matrices along any oriented path from node $1$ to node $i$. In particular, notice that the coherence condition guarantees that such a definition is path-independent: given two oriented paths $\gamma$ and $\gamma'$ from node $1$ to node $i$, their 
concatenation forms an oriented cycle, and coherence forces the corresponding matrix products to coincide. We then introduce the block-diagonal matrix $\mathcal S=\text{diag}(\mathbf O_{11},\mathbf O_{12},\dots,\mathbf O_{1N})$, where, without loss of generality, $\mathbf O_{11}\equiv\mathbf I_d$.

We have expressed the coherence condition through the product of transformation matrices along an oriented cycle. However it can be proved~\cite{tian2025matrix} to be equivalent to show that the vector $\vec{U}=\mathcal{S}^\top (\vec{1}_N\otimes \vec{v})$ satisfies $\mathcal{L}\vec{U}=0$, for any $\vec{v}\in\mathbb{R}^d$ and $\vec{1}_N=(1,\dots,1)^\top \in\mathbb{R}^N$. Interestingly, this observation allows us to establish the existence of the synchronous manifold by choosing $\vec{v}=\vec{s}(t)$, where $\vec{s}(t)$ solves~\eqref{eq:ODEs}. This again highlights the tight interplay between the topology of the MWN and the resulting dynamics.

Let us thus define
\begin{align}
    \label{eq:ZMWN}
    Z_i(\vec{x})=\frac{K}{N}\sum_j w_{ij}\mathbf{R}_{ij}\vec{x}_j\, ,
\end{align}
where $K>0$ is the coupling strength and $N$ the number of nodes. In the latter definition, the scalar weight $w_{ij}$ replaces the adjacency matrix of Eq.~\eqref{eq:defZ}, and moreover we introduce a generic orthogonal matrix depending on the indices $i$ and $j$, generalizing thus the models defined in~\cite{de2023generalized,buzanello2022matrix}. Since $\mathbf{R}_{ij}$ is an isometry, $\mathbf{R}_{ij}\vec{x}_j$ also lies on the sphere. Consequently, the model resembles the local mean-field term in~\eqref{eq:defZ}, with the key difference that node~$i$ does not directly perceive the state of its neighbour~$j$, but rather the transformed state obtained by mapping $\vec{x}_j$ through the edge.

The dynamical system describing the higher-dimension Kuramoto model coupled via a MWN thus reads
\begin{align}
    \label{eq:KuramMWN}
    \frac{d{\vec{x}}_i}{dt}= \Omega\vec{x}_i+Z_i(\vec{x})-\langle Z_i(\vec{x}),\vec{x}_i\rangle \vec{x}_i\, ,
\end{align}
$i=1,\dots,N$. Note that, by following the argument of the previous section, one can show that the flow of~\eqref{eq:KuramMWN} leaves the $(d-1)$--sphere invariant (and hence preserves the constraint $||\vec{x}_i||=1$ for all $i$).

Let us observe that by adapting the computation proposed in~\cite{de2023generalized}, the model~\eqref{eq:KuramMWN} returns a frustrated Kuramoto model where each link introduces its own lag. More precisely in the case $d=2$, the obtained model is
\begin{equation*}
    \dot{\theta}_i=\omega+\frac{1}{N}\sum_j w_{ij}\sin(\theta_j-\theta_i+\alpha_{ij})\, ,
\end{equation*}
where the lags $\alpha_{ij}$ are related to the weight matrices by
\begin{equation*}
    \mathbf{R}_{ij}=\left( \begin{matrix}\cos(\alpha_{ij}) & -\sin(\alpha_{ij})\\
    \sin(\alpha_{ij}) & \cos(\alpha_{ij}) \end{matrix} \right)\, .
\end{equation*}

\subsection{Global synchronisation}

To simplify equation~\eqref{eq:KuramMWN}, we perform the change of variables $\vec{y}_i=\mathbf{O}_{1i}\vec{x}_i$, or by using the stack vectors, $\vec{y}=\mathcal{S}\vec{x}$, then the latter rewrites
    \begin{align}
    \label{eq:KuramMWNSx}
    \frac{d{\vec{y}}_i}{dt}= \mathbf{O}_{1i}\Omega\mathbf{O}_{1i}^\top\vec{y}_i+\mathbf{O}_{1i}Z_i(\mathcal{S}^\top\vec{y})-\langle Z_i(\mathcal{S}^\top\vec{y}),\mathbf{O}_{1i}^\top\vec{y}_i\rangle \vec{y}_i\, .
\end{align}
By direct computation we get
\begin{align}
    \mathbf{O}_{1i}Z_i(\mathcal{S}^\top\vec{y})&=\frac{K}{N}\sum_j w_{ij}\mathbf{O}_{1i}\mathbf{R}_{ij}\mathbf{O}_{1j}^\top \vec{y}_j\notag\\
    &=\frac{K}{N}\sum_j w_{ij} \vec{y}_j\, ,
\end{align}
where we used the coherence condition to set $\mathbf{O}_{1i}\mathbf{R}_{ij}\mathbf{O}_{1j}^\top=\mathbf{I}_d$, indeed the latter product corresponds to an oriented cycle starting at node $1$ up to node $i$, from there jump to node $j$ and then back to node $j$. The term involving the scalar product returns
\begin{eqnarray}
    \langle Z_i(\mathcal{S}^\top\vec{y}),\mathbf{O}_{1i}^\top\vec{y}_i\rangle&=&\frac{K}{N}\sum_j w_{ij}\langle \mathbf{R}_{ij}\mathbf{O}_{1j}^\top\vec{y}_j,\mathbf{O}_{1i}^\top\vec{y}_i\rangle\notag\\
    &=&\frac{K}{N}\sum_j w_{ij}\langle \mathbf{O}_{1i} \mathbf{R}_{ij}\mathbf{O}_{1j}^\top\vec{y}_j,\vec{y}_i\rangle\notag\\
    &=&\frac{K}{N}\sum_j w_{ij}\langle\vec{y}_j,\vec{y}_i\rangle \vec{y}_i\, .
\end{eqnarray}
Let us moreover assume 
\begin{equation}
    \label{eq:condOmegai}
\mathbf{O}_{1i}\Omega\mathbf{O}_{1i}^\top=\Omega\, ,
\end{equation}
hence, in conclusion, we get 
\begin{align}
\label{eq:dKyvariab}    \frac{d{\vec{y}}_i}{dt}=\Omega \vec{y}_i+\zeta_i(\vec{y})-\langle \zeta_i(\vec{y}),\vec{y}_i\rangle \vec{y}_i\, ,
\end{align}
where we defined
\begin{align}
    \zeta_i(\vec{y}):=\frac{K}{N}\sum_j w_{ij}\vec{y}_j\, .
\end{align}
Namely, in the variable $\vec{y}_i$ we completely get rid of the rotation matrices and have thus mapped the $d$--Kuramoto model defined on MWN to a weighted scalar network.

We can then look for the existence of the synchronous solution $\vec{y}_i(t)=\vec{s}(t)$ for all $i=1,\dots, N$. By substituting the latter into~\eqref{eq:dKyvariab} we can prove that $\vec{s}$ solves
\begin{align}
    \label{eq:ODEs}
    \frac{d{\vec{s}}}{dt}=\Omega \vec{s}\, .
\end{align}
In conclusion we have proved the existence of a synchronous solution $\vec{y}_i(t)=\vec{s}(t)$ for all $i=1,\dots,N$, provided that $\mathbf{O}_{1i}\Omega=\Omega\mathbf{O}_{1i}$ for all $i$ and that the MWN is coherent. The latter property has been used to remove the dependence on the rotations $\mathbf{R}_{ij}$
 from Eq.~\eqref{eq:KuramMWN} so to obtain~\eqref{eq:dKyvariab}. In the next section we will show that coherence is also necessary to the existence of the synchronous manifold, requirement that can translated into the existence of the eigenvector $\mathcal{S}^\top (\vec{1}_n\otimes \vec{u})$ with zero eigenvalue for a suitable Laplace matrix.

% Let us now assume that, for all $i,j\in\{1,\dots,N\}$, there exists a vector $\vec{u}\in\mathbb{R}^d$ such that $\mathbf{R}_{ij}\vec{u}=\vec{u}$, namely all the rotation matrices preserve a given direction $\vec{u}$. Then, we can show that
% $\vec{s}(t)=\vec{u}s(t)$ is a spatially homogeneous solution, $\vec{x}_i(t)=\vec{s}(t)$ for all $i=1,\dots,N$, and satisfies
% \begin{align}
%     \label{eq:ODEs}
%     \dot{\vec{s}}=\Omega \vec{s}\, .
% \end{align}

% First we show that the term $Z_i(\vec{x})-\langle Z_i(\vec{x}),\vec{x}_i\rangle \vec{x}_i$ vanishes once evaluated on such function. We observe that
% \begin{align}
%     Z_i(\vec{s})=\frac{K}{N}\sum_j w_{ij}\mathbf{R}_{ij}\vec{s}=\frac{K}{N}\sum_j w_{ij}\vec{s}=\frac{K}{N}k_i\vec{s}\, ,
% \end{align}
% where we introduced the node strength $k_i=\sum_j w_{ij}$ and we used $\mathbf{R}_{ij}\vec{u}=\vec{u}$. Hence,
% \begin{align}
%     Z_i(\vec{s})-\langle Z_i(\vec{s}),\vec{s}\rangle \vec{s}=\frac{K}{N}k_i\vec{s}-k_i\frac{K}{N}\langle \vec{s},\vec{s}\rangle \vec{s} = 0\, ,
% \end{align}
% because $||\vec{s}||=1$. We can thus conclude that $\vec{x}_i=\vec{s}$ solves Eq.~\eqref{eq:KuramMWN}, because of the latter result and the fact that $\vec{s}$ solves~\eqref{eq:ODEs}.

%\subsection{Linear stability}
The system converges to complete synchronization if the solution $\vec{y}_i=\vec{s}$ is stable. To prove this claim, we will perform a linear stability analysis of~\eqref{eq:dKyvariab}    close to the latter solution by introducing $\vec{y}_i=\vec{s}+\delta \vec{y}_i$. Because $\vec{y}_i$ and $\vec{s}$ both lie on the $(d-1)$--sphere, we then have $\langle \delta\vec{y}_i,\vec{s}\rangle=0$ for all $i=1,\dots,N$.

Let us develop to first order Eq.~\eqref{eq:dKyvariab}   to obtain
\begin{widetext}
\begin{align}
    \label{eq:linKMWN}
    \frac{d{\delta\vec{y}}_i}{dt}=\Omega \delta\vec{y}_i+\frac{K}{N}\sum_j w_{ij}\delta\vec{y}_j-\frac{K}{N}\sum_j w_{ij}\delta\vec{y}_i-\frac{K}{N}\sum_j w_{ij}\langle \vec{s},\delta\vec{y}_i\rangle \vec{s}-\frac{K}{N}\sum_j w_{ij}\langle \delta\vec{y}_i,\vec{s}\rangle \vec{s}\, .
\end{align}
\end{widetext}
Notice that the two rightmost terms in the right hand side vanish because $\langle \delta\vec{y}_i,\vec{s}\rangle=0$, hence we can conclude that the time evolution of the perturbation $\delta\vec{y}_i$ is given by
\begin{align}
    \label{eq:linKMWN2}
    \frac{d{\delta\vec{y}}_i}{dt}&=\Omega \delta\vec{x}_i+\frac{K}{N}\sum_j w_{ij}\delta\vec{y}_j-\frac{K}{N}\sum_j w_{ij}\delta\vec{y}_i\nonumber\\
    &=\Omega \delta\vec{y}_i-\frac{K}{N}\sum_j \bar{\mathcal{L}}_{ij}\delta\vec{y}_j\, ,
\end{align}
where we introduced the (modified) {\em supra-Laplace matrix} $\bar{\mathcal{L}}=\bar{\mathbf{L}}\otimes \mathbf{I}_d$, where $\bar{\mathbf{L}}$ is the Laplace matrix of the underlying (scalar) weighted network, i.e., $\bar{L}_{ij}=k_i\delta_{ij}-w_{ij}$, with $k_i=\sum_j w_{ij}$. Let us observe that in the framework of MWN one can define the {\em supra-Laplace} matrix $\mathcal{L}$ whose $(i,j)$--entry is  given by
\begin{align}
    \label{eq:supL}
    \mathcal{L}_{ij} = \sum_t {w}_{it}\mathbf{I}_d  - \mathbf{W}_{ij}=k_i\mathbf{I}_d  - w_{ij}\mathbf{R}_{ij}\, ,
\end{align}
and one can prove~\cite{tian2025matrix} that the two are related by $\bar{\mathcal{L}}=\mathcal{S}\mathcal{L}\mathcal{S}^\top$.

To reduce the dimensionality of equation~\eqref{eq:linKMWN2} we will use the eigenbasis of the Laplace matrix, namely $\bar{\mathbf{L}}\phi^{(\alpha)}=\Lambda^{(\alpha)}\phi^{(\alpha)}$. More precisely, we project $\delta\vec{y}_i$ onto this basis:
\begin{align}
    \label{eq:proj}
    \delta\vec{y}_i=\sum_\alpha \delta\hat{y}_\alpha \phi^{(\alpha)}_i\, ,
\end{align}
to obtain from~\eqref{eq:linKMWN2}
\begin{align}
    \label{eq:linKMWN2yproj}
    \frac{d{\delta\hat{y}}_{\alpha}}{dt}=\left[\Omega -\frac{K}{N}\Lambda^{(\alpha)}\mathbf{I}_d \right]\delta\hat{y}_\alpha\, .
\end{align}

We are thus now dealing with a $1$--parameter family of linear systems, parametrized by $\Lambda^{(\alpha)}$. Let us define for all $\alpha$
\begin{align}
    \label{eq:MSF}
    \lambda(\Lambda^{(\alpha)})=\max_{j=1,\dots,d}\Re\left(\lambda_j(\Lambda^{(\alpha)})\right)\, ,
\end{align}
where $\lambda_j(\Lambda^{(\alpha)})$ are the $d$ roots of the characteristic equation associated to~\eqref{eq:linKMWN2yproj}, seen as functions of the eigenvalues $\Lambda^{(\alpha)}$. Being the matrix $\Omega$ antisymmetric, its eigenvalues are pure complex conjugate, or zero, hence $\lambda(\Lambda^{(1)})=0$ and $\lambda(\Lambda^{(\alpha)})=-\frac{K}{N}\Lambda^{(\alpha)}<0$, if $2\leq \alpha \leq N$.

We can thus conclude that synchronization emerges because $\lambda(\Lambda^{(\alpha)})<0$ for all $\alpha>1$, indeed under this assumption $\delta\hat{y}_\alpha(t)\rightarrow 0$ for $\alpha >1$, while $\delta\hat{y}_1(t)=e^{\Omega t}\delta\hat{y}_1(0)$. By using the projection on the eigenbasis and the form of $\vec{\phi}^{(1)}$, we can write $\delta\vec{y}_i(t)\sim e^{\Omega t} \sum_j \delta\vec{y}_j(0)/N$, in the limit of large $t$, and thus $\vec{y}_i(t)\sim \vec{s}(t)+e^{\Omega t} \sum_j \delta\vec{y}_j(0)/N$.
Let us however observe that in the original variables we get $\vec{x}_i(t)\sim \mathbf{O}_{1i}^\top\vec{s}(t)+ \mathbf{O}_{1i}^\top \sum_j \delta\vec{x}_j(0)/N$, namely we do not observe synchronization to $\vec{s}(t)$ in the $\vec{x}_i$ variables because of the presence of the transformations.

{In Fig.~\ref{fig:sync_y} we report the result of a numerical simulation in the case of $N=50$, $3$-Kuramoto oscillators anchored to the $2$-sphere and coupled via a MWN build by using the recipe provided in~\cite{gallo2026turing}. The underlying network is an Erd\H{o}s-R\'enyi graph with probability $p_{ER}=1/2$ to have a link. The matrix weights are rotations about the vertical axis, $\vec{u}=(0,0,1)^\top$, and the matrix $\Omega$ is the generator of the rotation about the same axis. The synchronous solution is $\vec{s}(t)=e^{\Omega t}\vec{s}(0)$, with $\vec{s}(0)=(1,1,1)^\top\sqrt{3}/3$. For random initial conditions (black dots in panel (a)) about $\vec{s}(0)$ (red dot), the system converges to the synchronous solution, i.e., $\vec{y}_i(t)\rightarrow \vec{s}(t)$ as shown by the blue dot superposed to the red one in panel (b). This claim is confirmed by the time evolution of the order parameter
\begin{align}
    \label{eq:ordpary}
    R_y(t)=\frac{1}{N}\Big|\Big|\sum_j \vec{y}_j(t)\Big|\Big|\, ,
\end{align}
that very quickly converges to $1$ (panel (c)).
\begin{figure*}[!ht]
\centering
\includegraphics[width=\textwidth]{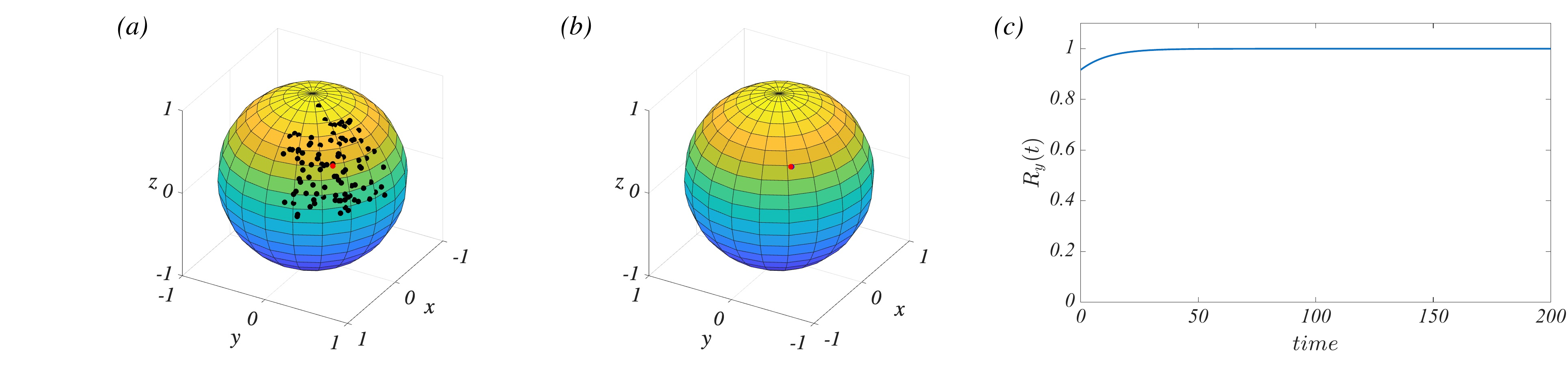}
\caption{\textbf{Dynamics for the $d$-Kuramoto model in the ``rotated'' variables, $\vec{y}_i$}. We report the initial conditions, $\vec{y}_i(0)$, (black dots in panel (a)) and the  configuration at some future time $t_{fin}$, i.e., $\vec{y}_i(t_{fin})$, (blue dot in (panel (b)) of the solution of Eq.~\eqref{eq:dKyvariab} in the case $d=3$, i.e., on the $2$--sphere. The MWN is build by using an Erd\H{o}s-R\'enyi network with $N=50$ nodes and probability $p_{ER}=0.5$ to have a link among a couple of nodes; the matrices $\mathbf{R}_{ij}$ are rotations about the vector $\vec{u}=(0,0,1)^\top$ and the matrix $\Omega$ is the generator of a rotation about the same vector. The synchronous solution (red dot in both panels) is given by $\vec{s}(t)=e^{\Omega t}\vec{s}(0)$, with $\vec{s}(0)=(\sqrt{3}/3,\sqrt{3}/3,\sqrt{3}/3)^\top$. Being $K=0.1$ the system synchronizes as we can appreciate by looking at the order parameter, $R_y(t)$ (panel (c)) that reaches quite fast the value $1$.}
\label{fig:sync_y}
\end{figure*}

Let us observe that the very same simulation presented in the original variables, $\vec{x}_i(t)$, cannot conclude about the emergence of synchronization (see Fig.~\ref{fig:sync_x}) where we report the initial, $\vec{x}_i(0)$ (panel (a)), and later time position, $\vec{x}_i(t_{fin})$ (panel (b)). In particular the latter panel clearly shows that $\vec{x}_i(t_{fin})$ distribute along a parallel, i.e., orthogonal to the rotation direction $\vec{u}=(0,0,1)^\top$; this claim is confirmed by observing that the order parameter $R_x(t)$
\begin{align}
    \label{eq:ordparx}
    R_x(t)=\frac{1}{N}\Big|\Big|\sum_j \vec{x}_j(t)\Big|\Big|\, ,
\end{align}
remains bounded away from $1$ (panel (c)).
\begin{figure*}[!ht]
\centering
\includegraphics[width=\textwidth]{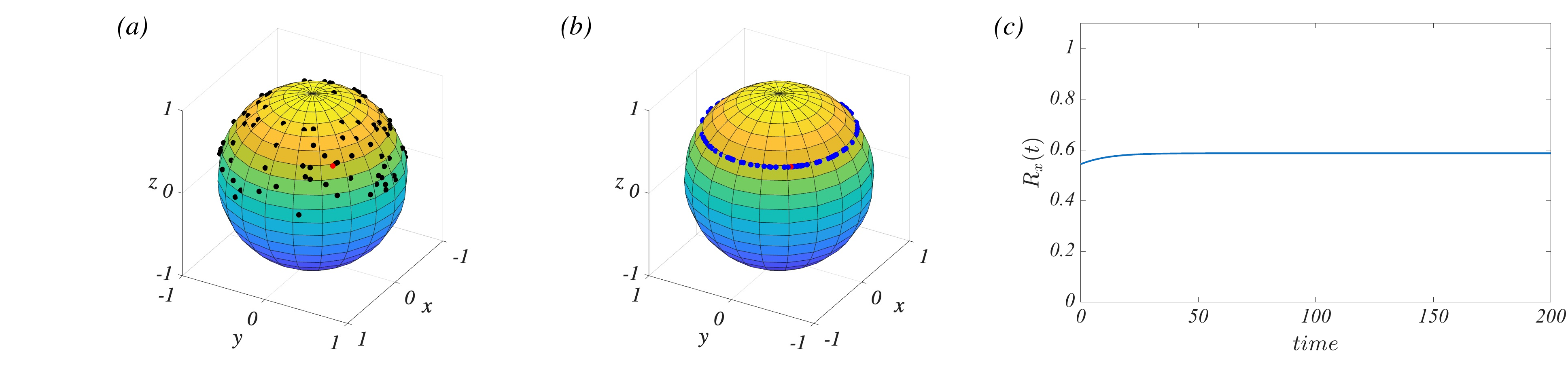}
\caption{\textbf{Dynamics for the $d$-Kuramoto model in the original variables, $\vec{x}_i$}. We report the same simulation shown in Fig.~\ref{fig:sync_y} but in the original variables, $\vec{x}_i$, instead of the rotated ones. Panel (a) shows the initial conditions, $\vec{x}_i(0)=\mathbf{O}_{1i}^\top \vec{y}_i(0)$, (black dots), while in panel (b) we report $\vec{x}_i(t_{fin})=\mathbf{O}_{1i}^\top \vec{y}_i(t_{fin})$ (blue dots). Even if the system synchronizes because $K=0.1$, this behaviour cannot be appreciated by using those coordinates as it is confirmed by looking at the order parameter, $R_x(t)$, (panel (c)) that does not grow beyond the value $\sim 0.6$.}
\label{fig:sync_x}
\end{figure*}}

Let us observe that the condition~\eqref{eq:condOmegai} is the very same condition imposed to ensure the existence of synchronization of oscillators coupled with a MWN~\cite{gallo2025global}. However, by relaxing this assumption and by looking at equation~\eqref{eq:KuramMWNSx} expressing the dynamics in the ``rotated'' variables $\vec{y}_i$, one can realise that the condition to obtain~\eqref{eq:dKyvariab} is $\mathbf{O}_{1i}\Omega_i\mathbf{O}_{1i}^\top=\Omega$. This last relation can be restated as follows: one may allow heterogeneous matrices $\Omega_i$, provided there exists an antisymmetric matrix $\Omega$ such that for all $i=1,\dots,N$ we have $\Omega_i=\mathbf{O}_{1i}^\top \Omega\mathbf{O}_{1i}$. Moreover, because $\mathbf{O}_{1i}=\mathbf{I}_d$, we have $\Omega_1=\Omega$. This means that the heterogeneity of the matrices can be ``reabsorbed'' into the transformations.

\begin{figure*}[!ht]
\centering
\includegraphics[width=\textwidth]{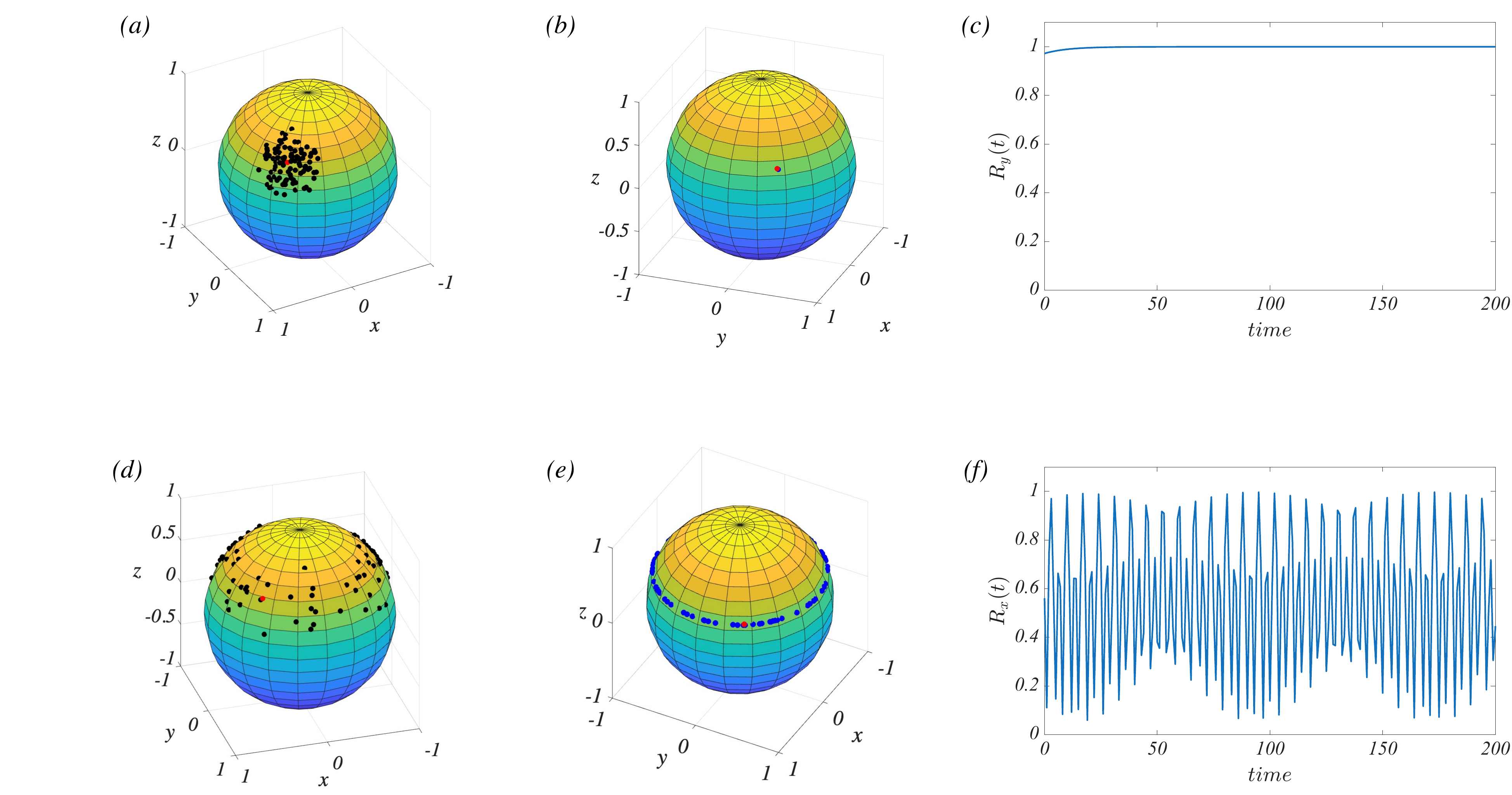}
\caption{\textbf{Dynamics for the $d$-Kuramoto model with $\Omega_i = \mathbf{O}_{1i}^\top \Omega\mathbf{O}_{1i}$}. We report a simulation for the case $K=0.1$, node dependent $\Omega_i$ matrices and matrices $\mathbf{R}_{ij}$ to be rotations about the axis $\vec{u}=(0,0,1)^\top$. Top panels refer to the ``rotated variables'', $\vec{y}_i$, while bottom panels to the original $\vec{x}_i$ ones. One can observe that starting from initial conditions, $\vec{y}_i(0)$, (black dots in panel (a)), randomly distributed about $\vec{s}(0)$ they eventually evolve synchronously to $\vec{s}(t)$ (see blue and red dots in panel (b)). Panel (c) show the order parameter $R_y(t)$ and it confirms the emergence of synchronization. The same time series, when plotted in the original variables, does not allow one to draw any conclusion about the presence of synchronization. Indeed random initial conditions (black points in panel (d)) will distribute on a sphere parallel (blue points in panel (e)) and the order parameter, $R_x(t)$ oscillates in time indicating the absence of synchronization. The simulation refers to $N=50$, $3$-Kuramoto oscillators coupled with a MWN built by using an Erd\H{o}s-R\'enyi network with probability to have a link given by $p_{ER}=0.5$. The matrix $\Omega$ is a random antisymmetric $3\times 3$ matrix. The synchronous solution (red dot in panels (a), (b), (d) and (e)) is given by $\vec{s}(t)=e^{\Omega t}\vec{s}(0)$, with $\vec{s}(0)=(\sqrt{3}/3,\sqrt{3}/3,\sqrt{3}/3)^\top$.}
\label{fig:sync_yOmegai}
\end{figure*}

It is worth stressing that the heterogeneity allowed by this construction is \textit{geometric} rather than spectral: since $\mathbf{O}_{1i}$ is orthogonal, all matrices $\Omega_i$ have the same spectrum, sharing the same eigenvalues as $\Omega$ and hence the same intrinsic rotation frequencies. Thus, the allowed heterogeneity consists solely in the orientation of the rotation axes, encoded in the transformation matrices $\mathbf{O}_{1i}$. In Fig.~\ref{fig:sync_yOmegai}, we report the results of numerical simulations for this case. We observe that, in the ``rotated variables'', the system exhibits synchronization, initial conditions (panel (a)) converge to the reference solution $\vec{s}(t)$ (panel(b)) and the order parameter, $R_y(t)$, converges to $1$ (panel (c)). On the other hand, original variables, $\vec{x}_i$, cannot permit to observe synchronization because of the presence of heterogeneous antisymmetric matrices $\Omega_i$.

\begin{figure*}[!ht]
\centering
\includegraphics[width=\textwidth]{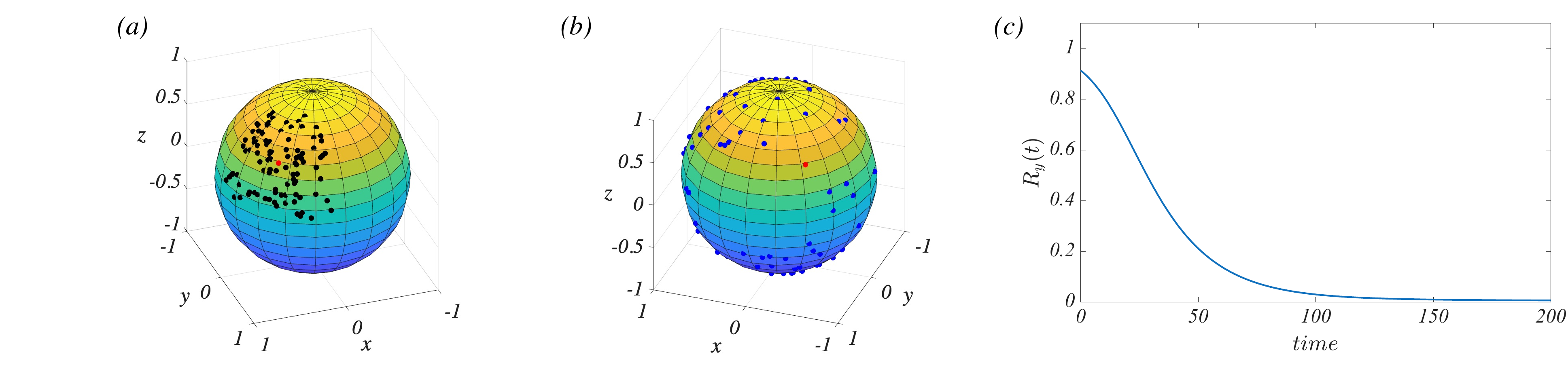}
\caption{\textbf{Dynamics for the $d$-Kuramoto model in the ``rotated'' variables, $\vec{y}_i$}. We report a simulation for the case $K=-0.1$. Initial conditions, $\vec{y}_i(0)$, (black dots in panel (a)) are randomly distributed about $\vec{s}(0)$; because $K<0$, the vectors $\vec{y}_i(t)$ cannot synchronize and are dispersed on the sphere (blue points in panel (b)). The order parameter, $R_y(t)$ converges to zero, thus testifying the absence of synchronization. The simulation refers to $N=50$, $3$-Kuramoto oscillators coupled with a MWN is build by using an Erd\H{o}s-R\'enyi network with probability to have a link given by $p_{ER}=0.5$. The matrices $\mathbf{R}_{ij}$ are rotations about the vector $\vec{u}=(0,0,1)^\top$ and the matrix $\Omega$ is the generator of a rotation about the same vector. The synchronous solution (red dot in both panels) is given by $\vec{s}(t)=e^{\Omega t}\vec{s}(0)$, with $\vec{s}(0)=(\sqrt{3}/3,\sqrt{3}/3,\sqrt{3}/3)^\top$.}
\label{fig:nosync_y}
\end{figure*}

To conclude, let us observe that the theory above defined applies to the case $K<0$ as well, the only difference being that the system can never synchronize because $\lambda(\Lambda^{(\alpha)})=-\frac{K}{N}\Lambda^{(\alpha)}>0$.
This claim can be appreciated by looking at the results presented in Fig.~\ref{fig:nosync_y}. Starting from initial conditions randomly distributed about $\vec{s}(0)$ (black points in panel (a)), the system is not capable to synchronize and indeed the variables $\vec{y}_i(t)$ do not align with $\vec{s}(t)$ (blue points in panel (b)). This is confirmed by the order parameter $R_y(t)$ that steadily decreases to zero (see panel (c)).

\section{Conclusion and perspectives}
\label{sec:conc}

The paper introduces a $d$--dimensional generalization of the Kuramoto model, where $N$ oscillators, coupled via complex networks or matrix-weighted networks (MWNs)~\cite{tian2025matrix}, evolve on the unit $(d-1)$--sphere. To any connected pair of nodes $i,j$ of the MWN, we associate a weight matrix $\mathbf W_{ij}:=w_{ij}\mathbf R_{ij}\in\mathbb R^{d\times d}$, where $w_{ij}\in\mathbb R$ and $\mathbf R_{ij}$ is an orthonormal transformation matrix. 
Our findings highlight a strong interplay between network topology and geometry of the interaction matrices. They rely on the condition that the MWN is coherent and moreover that the antisymmetric matrix $\Omega$ commutes with all rotation matrices $\mathbf{R}_{ij}$.  Our proof is based on the observation that the change of variables $\vec{y}_i=\mathbf O_{1i}\vec{x}_i$ leads to the removal of the explicit dependence on $\mathbf R_{ij}$ and that the synchronous solution $\vec{y}_i = \vec{s}$ satisfies $\frac{d{\vec{s}}}{dt}=\Omega\vec{s}$ provided $\Omega$ commutes with all $\mathbf O_{1i}$.

We then performed a linear stability analysis to show that, if the underlying network is connected, for any positive coupling strength, namely $K>0$, we always end up with a stable synchronous solution. This follows from the fact that the antisymmetric matrix $\Omega$ has purely imaginary eigenvalues, implying that the real part of the eigenvalues of the matrix $\Omega - \frac{K}{N}\Lambda^{(\alpha)}\mathbf{I}_d$ reads $-\frac{K}{N}\Lambda^{(\alpha)}$, which is strictly negative for all non-trivial modes $\alpha>1$. % Therefore, we conclude that, in contrast to classical scalar Kuramoto models with heterogeneous frequencies, synchronization does not require a critical coupling threshold in the identical-frequency setting considered here.
Therefore, we conclude that in the identical-frequency setting considered here, synchronization is guaranteed for any positive coupling strength $K>0$, provided the network is connected and the coherence condition is satisfied, as there is no competition between coupling and frequency mismatch and hence no critical threshold arises.

Several open problems naturally emerge from this work. The \emph{coherence} condition plays a central geometric role: it requires that the product of rotation matrices along any oriented cycle equals the identity, thereby preventing rotational mismatches across the network. Coherence is also the key property that enables a global change of variables reducing a matrix-weighted network to an equivalent scalar-weighted one. When this compatibility fails, such a reduction is no longer available and genuine geometric frustration may arise.
A natural direction is therefore to relax coherence and characterise the resulting dynamics. In incoherent networks one may expect frustrated synchronization patterns, by including cluster synchronization or other partially synchronized states, but a systematic analysis is still missing. Finally, we have assumed that the antisymmetric matrices belong to the same conjugacy class. Allowing genuinely heterogeneous intrinsic dynamics may lead to non-trivial synchronization thresholds, multistability, or bifurcation phenomena, and remains an interesting topic for future work.

Overall, the present study contributes to bridging higher-dimensional Kuramoto models and the theory of matrix-weighted networks, highlighting the interplay between network topology and geometric structure of the coupling. We expect that this framework may provide a useful starting point for the analysis of collective rotational dynamics in more general structured networks.

\section*{Author contributions}
A.G., R.L. and T.C. contributed equally to the project and the preparation of the manuscript.

\section*{Data Availability}
Data sharing is not applicable to this article as no datasets were generated or analyzed during the current study.

\section*{Code Availability}
The codes supporting the findings of this study are available from the authors upon request.

\section*{Competing interests}
The authors declare no competing interests.

\bibliography{mybib.bib}

\end{document}